\documentclass[useAMS]{mn2e}
\usepackage{aas_symbols}
\usepackage{times}
\usepackage{rotating}
\usepackage{multirow}
\long\def\symbolfootnote[#1]#2{\begingroup%
\def\thefootnote{\fnsymbol{footnote}}\footnote[#1]{#2}\endgroup}
\def\b{\beta}
\def\g{\gamma}
\def\D{\Delta}
\def\th{\theta}

\begin{document}

\title[Dust Scattering Model \& X-ray Plateaus]{The dust scattering model can not explain the shallow X-ray decay in GRB afterglows}

\author [Shen et al.]{R.-F. Shen$^{1}$\thanks{E-mail: rfshen@astro.as.utexas.edu (R-FS); rw@star.le.ac.uk (RW); pk@astro.as.utexas.edu (PK); pto@star.le.ac.uk (PTO); pae9@star.le.ac.uk (PAE)}, R. Willingale$^{2}$\footnotemark[1], P. Kumar$^{1}$\footnotemark[1], P. T. O'Brien$^{2}$\footnotemark[1] and P. A. Evans$^{2}$\footnotemark[1]\\
$^{1}$Department of Astronomy, University of Texas at Austin, Austin, TX 78712, USA\\
$^{2}$Department of Physics and Astronomy, University of Leicester, Leicester LE1 7RH, UK}

\date{Accepted 2008 November 10. Received 2008 November 9; in original form 2008 June 21}

\pagerange{\pageref{000}--\pageref{000}} \pubyear{2009}

\maketitle

\begin{abstract}
 A dust scattering model was recently proposed to explain the shallow X-ray decay (plateau) observed prevalently in Gamma-Ray Burst (GRB) early afterglows. In this model the plateau is the scattered prompt X-ray emission by the dust located close (about 10 to a few hundred pc) to the GRB site. In this paper we carefully investigate the model and find that the scattered emission undergoes strong spectral softening with time, due to the model's essential ingredient that harder X-ray photons have smaller scattering angle thus arrive earlier, while softer photons suffer larger angle scattering and arrive later. The model predicts a significant change, i.e., $\D \b \sim 2 - 3$, in the X-ray spectral index from the beginning of the plateau toward the end of the plateau, while the observed data shows close to zero softening during the plateau and the plateau-to-normal transition phase. The scattering model predicts a big difference between the harder X-ray light curve and the softer X-ray light curve, i.e., the plateau in harder X-rays ends much earlier than in softer X-rays. This feature is not seen in the data. The large scattering optical depths of the dust required by the model imply strong extinction in optical, $A_V \gtrsim $ 10, which contradicts current findings of $A_V= 0.1 - 0.7$ from optical and X-ray afterglow observations. We conclude that the dust scattering model can not explain the X-ray plateaus.
\end{abstract}

\begin{keywords}
dust - scattering - radiation mechanisms: non-thermal - gamma-rays: burst
\end{keywords}

%%%%%%%%%%%%%%%%%%%%%%%%%%%%%%%%%%%%%%%%%%%%%%%%%%%%%%%%%%%%%%%%%%%%%%%%%%%%%%%%
%%%%%%%%%%%%%%%%%%%%%%%%%%%%%%%%%%%%%%%%%%%%%%%%%%%%%%%%%%%%%%%%%%%%%%%%%%%%%%%

\section{Introduction}

{\it Swift} has discovered a generic behaviour in X-ray afterglows of Gamma-Ray Bursts (GRB): the X-ray light curve (LC) first shows a steep decline during a few hundred seconds after the end of the $\g$-rays, then it shows a shallow decay lasting $10^4 - 10^5$ s which is followed by a ``normal'' power-law decay (Nousek et al. 2006; O'Brien et al. 2006). The normal decay at late times is the canonical afterglow component due to the interaction of the decelerated GRB ejecta with the circumburst medium, i.e., the forward shock model. The steep decline is generally interpreted to have the same origin as the prompt $\g$-ray emission (e.g., Kumar \& Panaitescu 2000; Liang et al. 2006).

The intervening shallow decay, sometimes called the `plateau', is the most puzzling feature of the X-ray LC. The most straightforward interpretation is a late steady energy injection into the external shock, where the latter is produced by the decelerated early ejecta plunging into the medium. The late energy injection could be due to a new ejecta from the late activity of the central engine (e.g., Dai \& Lu 1998a,b; Zhang \& M\'{e}sz\'{a}ros 2001; Dai 2004; Yu \& Dai 2007), or due to a slow trailing part of the outflow catching up with the already forward-shock-decelerated early part of the outflow when the outflow has a spread in its Lorentz factor distribution (e.g., Granot \& Kumar 2006). If it is the first scenario, then this interpretation implies a steady, late activity of the central engine -- lasting as long as a day -- which poses a challenge to the models of the central engine. Moreover, according to the energy-injection interpretation, the plateau-to-normal transition in the LC corresponds to the cessation of the energy injection, thus the transition should be achromatic. But in about 1/3 of the X-ray plateau GRBs with optical afterglow observations, the optical LC does not show a simultaneous plateau-to-normal break, while in another smaller fraction of the plateau cases, the plateau-to-normal breaks in optical and X-ray are indeed simultaneous (Panaitescu 2007). In most cases the power-law decay following the plateau is consistent with the predictions (the closure relationships) of the forward shock model, which in turn is consistent with the energy injection interpretation. There is a long list of alternative models for the plateau phase, such as a slow energy transfer from the ejecta to the ambient medium (Kobayashi \& Zhang 2007), a two-component jet model (e.g., Granot et al. 2006), a varying shock microphysical parameter model (e.g., Panaitescu et al. 2006), and a reverse shock dominated afterglow model (Uhm \& Beloborodov 2007; Genet et al. 2007), etc. (see Zhang 2007 for a review), but none of them satisfy all the observational constraints.

An attractive possibility was suggested by Shao \& Dai (2007) regarding the origin of the X-ray plateau. If the long-duration GRB progenitors are massive stars, it is very likely that dust exists in the vicinity of the GRB site since it is in a star forming region. The X-ray photons from the GRB and its afterglow can be scattered in small angles by the dust near the line-of-sight to the GRB, as analogous to the halo emissions of other X-ray sources (e.g., Smith \& Dwek 1998). The GRB prompt emission scattered off the dust has been considered earlier by Esin \& Blandford (2000) and M\'{e}sz\'{a}ros \& Gruzinov (2000). Aside from the scattering by the dust local to the GRB site, Miralda-Escud\'{e} (1999) considered the scattering of the X-rays from the GRB afterglows by the dust in the intervening galaxies along the line-of-sight to the GRB, but the flux turns out to be very low and difficult to detect for that case. Depending on the distance of the local dust region to the GRB site, a delayed emission component from the scattering can show up in the afterglows. Shao \& Dai (2007) and Shao et al. (2008) recently used this scenario to interpret the plateau phase in the X-ray afterglow LC as to be the scattered prompt X-rays by the dust located at about ten to a few hundred pc from the GRB site. The scattering happens preferentially within a characteristic scattering angle $\th_c$ which is dependent on the photon energy $E$ and the dust grain size. At larger angles the differential scattering cross section of the dust grains decays steeply. Therefore the scattering within $\th_c$ gives rise to a plateau phase whose duration is determined by $\th_c$ and the distance of the dust region to the GRB site. Larger angle scattering produces a $F(t)\propto t^{-2}$ decay following the plateau. This model does not need to invoke a long steady central engine activity. In addition, since the scattering only works in the X-ray band, the lack of a simultaneous break in optical LC does not pose a problem for this model.

The purpose of this work is to carefully investigate the output of this dust scattering model - in terms of the spectral and temporal properties of the scattered emission - and to compare it with the data. The paper is structured as follows. We first calculate and quantify the softening expected from the dust scattering model in Section 2. Then, we search in the data for evidence in favour of the model including the spectral evolution in the plateau and post-plateau phases for a sample of GRBs in Section 3 and 4. An expected difference in hard X-ray and soft X-ray LCs is discussed in Section 5. We calculate and discuss the optical extinction for the dust in Section 6. Our conclusion and further discussion are presented in Section 7. Throughout the paper the spectral index $\b$ and the time decay index $\alpha$ of the emission flux are defined as in $f_{\nu}(t) \propto \nu^{-\b} t^{-\alpha}$.

%%%%%%%%%%%%%%%%%%%%%%%%%%%%%%%%%%%%%%%%%%%%%%%%%%%%%%%%%%%%%%%%%%%%%%%%%%%%%%%%%
\section{Spectral softening in dust echo emission}

We first derive the temporal and spectral properties of the scattered emission or the `echo' (hereafter we use `echo' and `scattered emission' interchangeably) by the dust in the simplest geometry where the dust is concentrated in a thin layer (or a dust ``screen'') near the GRB, following Shao \& Dai (2007). Then we consider a generalised geometry where the dust is distributed in an extended zone.

Let us consider a dust ``screen'' located at a distance $R$ from the GRB source. This dust screen does not have to enclose entirely the GRB source, as long as its angular size with respect to the GRB site is larger than the characteristic scattering angle $\th_c$ (see below). The grains in the dust have a size distribution $dN(a)/da \propto a^{-q}$ within a range ($a_-, a_+$), where $a$ is the grain size, $q$ is the distribution index and $N(a)$ is the column density of all grains with size $\le a$. In this paper we use these typical values $a_-$= 0.025 $\mu$m, $a_+$= 0.25 $\mu$m, and $q$= 3.5 inferred from the observations (Mathis et al. 1977; Mauche \& Gorenstein 1986; Draine 2003). We found that adopting other typical values did not change our main results. Consider a GRB source with a fluence per unit energy $S(E)$ [erg cm$^{-2}$ keV$^{-1}$] at X-ray photon energy $E$. Since the GRB source duration ($\sim$ 10 s) is much shorter than the plateau, it can be considered as being instantaneous.

The flux of the dust scattered emission per photon energy, per grain size, at time $t$ can be estimated by
\begin{equation}
 F_{E, a}(t)= \frac{S(E)}{t} \tau[E, a, \hat{\th}(t)],
\end{equation}
where $\tau[E, a, \hat{\th}(t)]$ is the scattering optical depth per grain size $a$, to the photon with energy $E$ and at the scattering angle $\hat{\th}(t)$; $\hat{\th}(t)$ is given by the geometrical relation $t= R \th^2/(2c)$.

The angular part of the optical depth can be separated out from $\tau$ by
\begin{equation}
\tau[E, a, \hat{\th}(t)]= 2\tau_a(E) j_1^2[\hat{x}(E, a, t)],
\end{equation}
where $\tau_a(E)$ is the total optical depth per grain size $a$ and to the photon energy $E$; $j_1(x)=\sin(x)/x^2-\cos(x)/x$ is the spherical Bessel function of the first order which describes the scattering-angle dependence of the cross section, and $\hat{x}= 2\pi E a \th /(hc)$ is the scaled scattering angle where $h$ is the plank constant and $c$ is the light speed (Overbeck 1965; Alcock \& Hatchett 1978). Via the geometrical relation, $\hat{x}$ can be expressed in terms of $E$, $a$ and $t$:
\begin{equation}
\hat{x}= \frac{2\pi}{hc}\sqrt{\frac{2ct}{R}} E a.
\end{equation}

$j_1^2(x)$ increases as $\propto x^2$ from $x=0$ to $x \simeq 1.5$ and then drops rapidly as $\propto x^{-2}$ for $x > 1.5$. Therefore, at a given photon energy $E$, the echo flux LC first appears as a plateau, then transitions to a decay as steep as $\propto t^{-2}$. The transition time, which corresponds to a characteristic scattering angle $\th_c$ and in turn to $\hat{x} \simeq 1.5$, would be given by
\begin{equation}
t_c= 4.5\times10^4 \biggl(\frac{E}{{\rm 1 keV}}\biggr)^{-2}\biggl(\frac{R}{100 {\rm pc}}\biggr)\biggl(\frac{a}{0.1 {\rm \mu m}}\biggr)^{-2}\,\, {\rm s}.
\end{equation}

We see from Eq. (4) that the duration of plateau is very sensitive to the photon energy: the plateau at higher energies ends much earlier than that at lower energies. Thus the overall echo emission must experience strong spectral softening. Note that if the echo is observed within a finite energy range, such as in the XRT band (0.3 - 10 keV), the softening must have begun long before the end of the plateau, because the overall plateau ending time is determined by $t_c$ of the softest photon while the softening begins at $t_c$ of the hardest photon; the ratio of the two times is the ratio of the photon energies reversed and squared, e.g., a factor of 1000 for the XRT band.

The dependence of $\tau_a(E)$ on energy and grain size is
\begin{equation}\label{tau_a_E}
\tau_a(E)= \tau_0(E=1 {\rm keV}, a= 0.1 {\rm \mu m}) (\frac{E}{1 {\rm keV}})^{-s} (\frac{a}{0.1 {\rm \mu m}})^{4-q};
\end{equation}
in the Rayleigh - Gans approximation, $s=$ 2 (van de Hulst 1957; Overbeck 1965; Hayakawa 1970; Alcock \& Hatchett 1978; Mauche \& Gorenstein 1986).

The echo emission spectrum at an observer time $t$ is obtained by
\begin{displaymath}
F_E(t)= \int_{a_-}^{a_+} F_{E,a}(t) da
\end{displaymath}
\begin{equation} \quad \quad \quad
\propto \frac{S(E)E^{-s}}{t} \int_{a_-}^{a_+} a^{4-q} j_1^2[\hat{x}(E,a,t)] da.
\end{equation}
The LC can be obtained by integrating $F_E(t)$ over a desired energy bandpass.

The softening can be seen from Eq. (6) as follows. Since $\hat{x} \propto t^{1/2} E a $, at some given time $t$, $\hat{x}$ might be $> 1.5$ for the hard photons and $< 1.5$ for the soft photons, while the intermediate photon energy that defines and separates the ``soft'' and the ``hard'' corresponds to $\hat{x} \approx 1.5$ and it decreases with time. For hard photons, $j_1^2(\hat{x}) \propto \hat{x}^{-2}$. Taking $E$ out of the integral in Eq. (6) gives $F_E(t) \propto S(E)E^{-s-2}$, so the spectral index is increased by 4 for $s=2$. For soft photons, $j_1^2(\hat{x}) \propto \hat{x}^2$, so $F_E(t) \propto S(E)E^{-s+2}$ and the spectral index is unchanged for $s=2$. Therefore, the softening happens first in the high energy part of the spectrum, and then propagates toward the lower energies with time, until the spectrum in the whole bandpass is softened - this is also when the plateau of the overall LC approaches to its end - with a change of the overall spectral index $\D\b = 4$ with respect to the source spectrum (cf. Fig. 2 below). This change in $\b$ should be easy to detect if the echo emission dominates the plateau.

We calculate the LCs of the dust echo in the XRT band for a variety of dust parameter values and the echo spectrum at different times. They are exactly the same as those obtained by Shao \& Dai (2007) in their Fig. 3 and Fig. 4 and thus confirm their results and the analytical scalings derived above.

%%%%%%%%%%%%%%%%%%%%%%%%%%%%%%%%%%%%%%%%%%%%%%%%%%%%%%%%%%%%%%%%%%%%%%%%%%%%%%%%%%%%%%%%%%%%%%%%%%

\subsection{Extended dust zone}

In the vicinity of GRBs, the dust zone may extend over
a large distance. To study the difference in the echo emission properties of an extended dust zone and of a thin dust layer, we consider in this subsection a power-law dust distribution over a distance range $[R_-, R_+]$ with the dust number density profile $n(R)\propto R^{-\delta}$, where $R$ is the distance to the source. The grain properties, e.g., size distribution, are assumed to be independent of $R$. For ease of calculation, we divide the extended dust zone into a series of $N$ discrete thin dust layers ($N \ge$ 30; a change of $N$ does not affect the results), located progressively further from the GRB with equal separation in the $\log(R)$ scale. The scattered flux and its spectrum at any given time is the sum of the contributions from all dust layers at that time.

Fig. 1 shows the LC of the scattered emission from an extended dust zone for varied sets of parameters. It can be seen that the ending time of the plateau is mainly determined by the location of the inner boundary of the dust zone. This is not surprising because the density of the dust is decreasing with radius thus the scattering LC arises mainly from the inner rim of the dust zone. Fig. 2 shows the spectra of the scattered emission at different times, from which the softening is evident. The LCs in Fig. 1 and the spectra in Fig. 2 are almost same as the ones for a single dust layer model (cf. the \emph{thin solid} line in Fig. 1 and Shao \& Dai (2007)'s results in their Fig. 3 and 4), which shows that the generalisation of the model to an extended dust zone does not change much the temporal or spectral behavior of the scattered emission.

We also calculate the overall spectral index in the XRT band at each observer time and the instantaneous decay index of the LC, which are shown in Fig. 3. Note that, due to the softening, the echo spectrum during the plateau is no longer a single power law function (see Fig. 2). Thus we calculate a ``pseudo'' spectral index $\b_{0.3-10}$ using the flux densities at the two ends of the XRT band, 0.3 keV and 10 keV, respectively, to illustrate the extent of softening with respect to the source spectrum.

The results show that the dust echo emission must experience significant spectral softening; the spectral slope increased by $\D\b \approx 3$ from the early phase of the plateau to the end of the plateau. A more realistic extended dust zone model brings no notable change to this property.

%%%%%%%%%%%%%%%%%%%%%%%%%%%%%%%%%%Begin Fig. 1%%%%%%%%%%%%%%%%%%%%%%%%%%%%%%%%%%%%%
\begin{figure}
\centerline{
\includegraphics[width=6.4cm,angle=270]{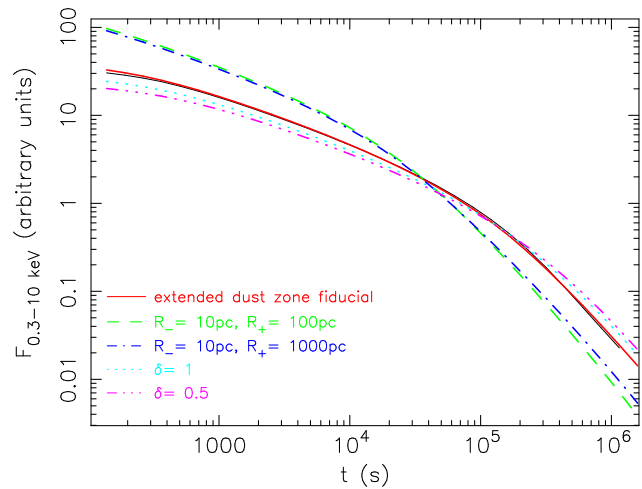}}
\caption{The flux light curve in the XRT band of the echo emission from an extended dust zone with a distance range $[R_-, R_+]$. The \emph{thick solid} line (\emph{red} in electronic version) is for the fiducial values used for the parameters of the dust spatial distribution model: $R_-$= 50 pc, $R_+$= 500 pc, $\delta$= 2. The parameter values for the dust grain properties (same for all the curves) are $a_-$= 0.025 $\mu$m, $a_+$= 0.25 $\mu$m, $q$= 3.5, and $s$= 2. The parameter values listed in the legends are the only ones that are changed each time. The assumed source spectrum is a flat power law with a high energy cut-off: $S(E)\propto E^0 \exp(-E/100 {\rm keV})$. For comparison, the LC from a single dust screen located at $R= 100$ pc with the same total optical depth is also shown here as the (\emph{black}) \emph{thin solid} line.}
\end{figure}
%%%%%%%%%%%%%%%%%%%%%%%%%%%%%%%%%%%%%End Fig. 1%%%%%%%%%%%%%%%%%%%%%%%%%%%%%%%%%%%

%%%%%%%%%%%%%%%%%%%%%%%%%%%%%%%%%%%Begin Fig. 2%%%%%%%%%%%%%%%%%%%%%%%%%%%%%%%%%%%%
\begin{figure}
\centerline{
\includegraphics[width=6.4cm,angle=270]{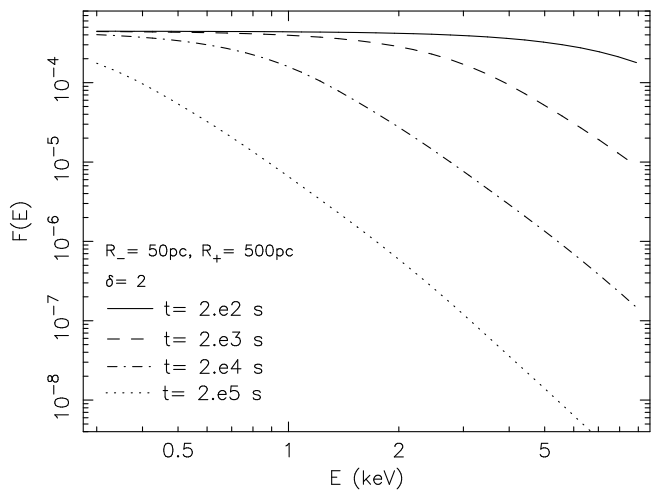}}
\caption{The spectrum of the echo emission from an extended dust zone at different observer times. The propagation of the softening toward the low energies is evident. The model parameter values are the same as the fiducial ones in Fig. 1. The assumed source spectrum is a flat power law with a high energy cut-off: $S(E)\propto E^0 \exp(-E/100 {\rm keV})$. The change of the spectral index due to the softening is found to be insensitive to the source spectral index.}
\end{figure}
%%%%%%%%%%%%%%%%%%%%%%%%%%%%%%%%%%%%End Fig. 2%%%%%%%%%%%%%%%%%%%%%%%%%%%%%%%%%%%%%

%%%%%%%%%%%%%%%%%%%%%%%%%%%%%%%%%%%Begin Fig. 3%%%%%%%%%%%%%%%%%%%%%%%%%%%%%%%%%%%%
\begin{figure}
\centerline{
\includegraphics[width=6.4cm,angle=270]{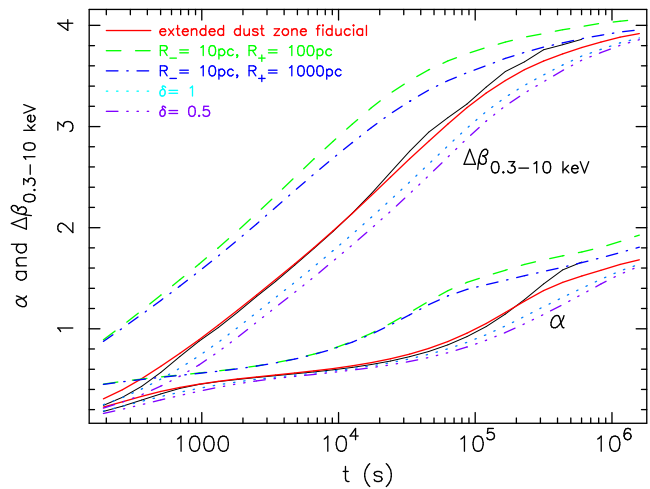}}
\caption{The instantaneous flux decay index $\alpha$ (defined as $F_{0.3-10 {\rm keV}}(t) \propto t^{-\alpha}$) and the spectral softening $\D \b = \b_{0.3-10 {\rm keV}} -\b_0$ for the echo emission from an extended dust zone, where $\b_{0.3-10 {\rm keV}}= \log[F_E(0.3 {\rm keV})/F_E(10 {\rm keV})]/\log(10/0.3)$ is the two-point spectral index and $\b_0$ is the source emission spectral index. In the cases plotted here we use $\b_0=0$. By changing the values for $\b_0$, e.g., to $\b_0$= 1, -0.5 or -1, we find that the calculated $\D \b$ is insensitive to $\b_0$. Each pair of lines for $\alpha$ and $\D\b$ of the same line style correspond to a same set of model parameter values. The \emph{thick solid} lines (\emph{red} in electronic version) are for the same fiducial model parameter values used in Fig. 1. For comparison, the $\alpha$ and $\D \b$ for the single dust screen model with the same parameter values as that in Fig. 1 are also shown as the \emph{thin solid} lines (\emph{black} in electronic version). }
\end{figure}
%%%%%%%%%%%%%%%%%%%%%%%%%%%%%%%%%%%%End Fig. 3%%%%%%%%%%%%%%%%%%%%%%%%%%%%%%%%%%%%%

%%%%%%%%%%%%%%%%%%%%%%%%%%%%%%%%%%%%%%%%%%%%%%%%%%%%%%%%%%%%%%%%%%%%%%%%%%%%%%%%%%%%%%%%

\section{Search for spectral evidences in the data}

In this section we describe our search for the statistical evidence in the X-ray data during the plateau that can support the dust scattering model. There are two pieces of evidence that we are looking for. First, if the plateau is due to the prompt X-rays scattering off the dust, in the early phase of the plateau when the spectral softening has not yet begun, the spectral index of the scattered emission must be the same as that of the prompt X-rays. Thus we expect to see in the data a correlation between the spectral index of the plateau, which we denote as $\b_a$ here, and that of the prompt X-rays. There are two complications to note. (1) For the prompt emission, usually the X-ray spectral index is unavailable so we have to use the one for the prompt $\g$-rays, $\b_{\g}$, to represent it; in some cases the X-ray slope might be shallower than the $\g$-ray slope by $1/2$ due to a cooling break. (2) The published spectral index for the plateau is usually measured from the photon counts integrated over the whole plateau duration, therefore this spectrum might be softer than at the beginning of the plateau. Nevertheless, a mild trend of the correlation in the data should still be expected.

The second evidence is based on the strong softening predicted by the model as was demonstrated in Fig. 2 - 3, which show a strong evolution of the spectral index during the plateau and until its end. Thus, if the model is correct, the distributions of the spectral index during the plateau, $\b_a$, for a sample should be significantly smaller than that measured in the post-plateau phase, denoted as $\b_{ad}$.

\subsection{Sample}

A sample of GRBs showing X-ray plateaus with sufficient spectral and temporal information is needed to check for these two evidences. Willingale et al. (2007) analysed 107 Swift XRT detected GRB afterglows and found 80\% of the bursts show a plateau in the X-ray LC. Out of the 80\% of total bursts sample, 54 have both spectral indices before and after the end of plateau available. We further reduced the sample down to 26 bursts; we rejected those bursts that had one of the following properties: (1) the temporal decay slope $\alpha >$0.8, too steep to be defined as a ``plateau''; (2) XRT coverage is very sparse or long gaps exist during the plateau; (3) the ``plateau'' is actually due to one or more flares. We have also included 24 bursts from the sample of Liang et al. (2007) that satisfy the above criteria.

%%%%%%%%%%%%%%%%%%%%%%%%%%%%Begin Fig. 4%%%%%%%%%%%%%%%%%%%%%%%%%%%%%%%%%%%%%%%%%%%%%%%
\begin{figure*}
\centerline{
\includegraphics[width=6.55cm,angle=270]{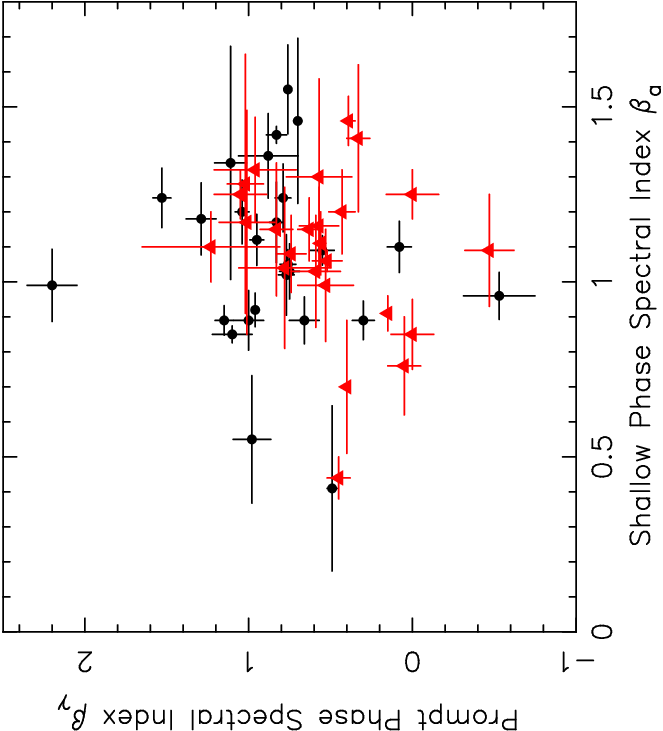}
\hspace{0.5cm}
\includegraphics[width=6.55cm, angle=270]{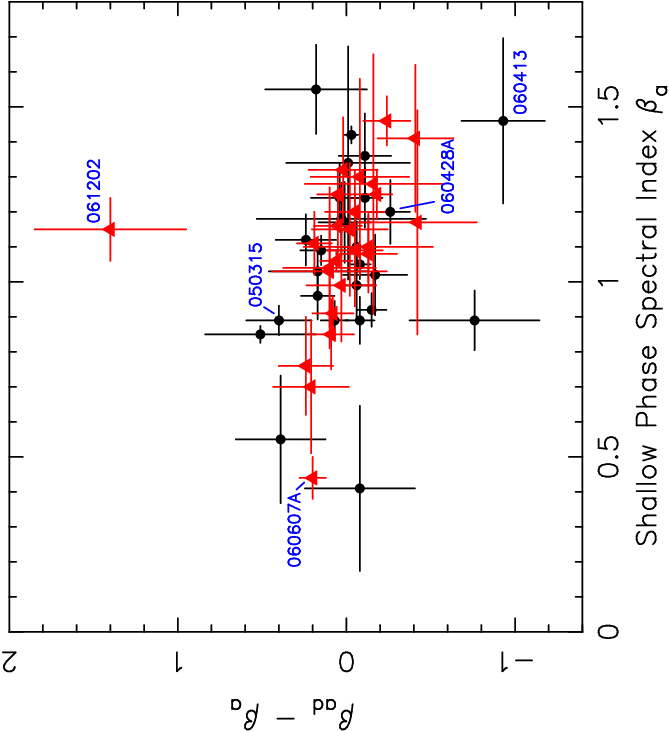}}
\caption{{\bf Left}: The prompt phase BAT spectral index $\b_{\g}$ vs. the shallow phase XRT spectral index $\b_a$. {\bf Right}: The change in the spectral index $\b_{ad} - \b_a$ vs. $\b_a$, where $\b_{ad}$ is the spectral index of the decay after the end of the shallow phase. Several individual cases of GRBs which show evidence of spectral softening or hardening are labeled. The spectral index $\b$ is defined as $f_{\nu} \propto \nu^{-\b}$. The error bars are 1-$\sigma$ errors. The sample is selected from Willingale et al. (2007) (\emph{filled circles}, and in \emph{black} color in electronic version) and Liang et al. (2007) (\emph{filled triangles}, and in \emph{red} color in electronic version).}
\end{figure*}
%%%%%%%%%%%%%%%%%%%%%%%%%%%%%%End Fig. 4%%%%%%%%%%%%%%%%%%%%%%%%%%%%%%%%%%%%%%%%

\subsection{Results}

We plot $\beta_{\g}$ vs. $\beta_a$ and the difference between $\b_{ad}$ and $\b_a$ vs. $\b_a$ in Fig. 4. No clear correlation between $\b_{\g}$ and $\b_a$ is seen, which disfavours the dust model. Generally $\b_a$ is softer than $\b_{\g}$. This is consistent with the expected softening during the plateau. However, if it was the softening that could have weakened or broken the expected correlation, there must be a bigger scatter in $\b_a$ than in $\b_{\g}$. The data shows the contrary: $\b_a$ is in the range of 0.5 - 1.5 where $\b_{\g}$ is in -0.5 - 1.5. Thus the comparison between $\b_a$ and $\b_{\g}$ is inconsistent with the model expectation.

Moreover, no dominant softening trend in the spectral index is seen from the plateau to the post-plateau phase; bursts with smaller $\b_a$ show slight softening and those with larger $\b_a$ show slight hardening. Most of the bursts show zero spectral change across the end of the plateau within 1-$\sigma$ measurement error. Only three bursts show evidence of spectral softening - two bursts at 2-$\sigma$ level (GRB 050315: $\D \b$= 0.4; GRB 060607A: $\D \b$= 0.2) and one burst at 3-$\sigma$ level (GRB 061202: $\D \b$= 1.4). There are also two bursts showing spectral hardening - one at 2-$\sigma$ level (GRB 060428A: $\D \b$= -0.26) and one at 3-$\sigma$ level (GRB 060413: $\D \b$= -0.93). Those individual cases are marked in the right panel of Fig. 4.

The two results -- no correlation between $\b_{\g}$ and $\b_a$ and no clear difference between $\b_a$ and $\b_{ad}$, also reported in Willingale et al. (2007) and Liang et al. (2007) -- are inconsistent with the expectations of the dust scattering model.

\section{Time history of the spectral during the plateau}

The model predicts a significant spectral evolution from the beginning ($\sim$ 200 s after the burst) to the end of the dust echo plateau. The spectral index shows a monotonic increase by $\D\b \approx 3 - 4$. If the X-ray plateau is indeed due to or dominated by the dust echo emission, the strong spectral softening can be very easily detected in the XRT data. To compare the data with this expectation, we look closely at the time resolved spectral information during the plateau for the best observed GRBs.

To determine whether there is any dominant trend of spectral evolution during the plateau phase, we compile a sample of 21 GRBs with well defined plateau phases, excellent time coverage and good signal-to-noise ratios. For details of how the X-ray light curves used in this work were produced, see Evans et al. (2007). This sample is listed in Table 1.

For each GRB, the overall XRT LC is considered to be composed of two components. The first one is the very rapid decay just following the $\g$-rays. The second component is the plateau and the subsequent normal power-law decay. Both components are well fitted by the same functional form as introduced by Willingale et al. (2007). We define $T_1$ -- the time of transition from the first to the second component -- as the time when the two components are equal; this is a good measure of the start of the plateau. $T_2$ is the end of the plateau and the start of the final power-law decay.

We plot the hardness ratio, as defined by the ratio of the photon counts in 1.5 - 10 keV and 0.3 - 1.5 keV bands, for each time interval of coverage during the plateau phase. We find that all the hardness ratio changes through the plateau phase are quite small and, for the bursts with the largest change the hardness ratios are getting harder (near the start of plateau), not softer. The ones that get softer do so only slightly. This confirms the findings by Butler \& Kocevski (2007).

We measure the spectral index taking into account photo-absorption at lower X-ray energies. We use the absorbed-power-law fit to the spectrum of the early XRT data - mainly the steep decline phase in the LC, which contains most of the photon counts - to determine the neutral $H$ column density given as a combination of two components - the Galactic column and a host intrinsic one. Then we use this neutral $H$ absorption model to convert the measured hardness ratio in the plateau into the spectral index with an appropriate error.

For many GRBs the time coverage during the plateau is rather patchy. Thus we select a time window which includes the plateau and takes into account the coverage. Sometimes it has to include data before $T_1$ and after $T_2$ so that the behaviour across the plateau is well constrained. The evolution of the spectral index over the selected time window is fitted as a linear function of the logarithmic time. Extrapolating the best fit function to both sides of the window gives the spectral indices $\b_1$ and $\b_2$ at $T_1$ and $T_2$, respectively. These are the best estimates of the spectral index at the start and the end of the plateau.

Fig. 5 shows a few examples of the $\b$-evolution during the plateau. The observed $\b$-evolutions for our sample are tabulated in Tab. 1.  None of the afterglows show a notable softening over the plateau. Most show zero evolution of $\b$, with very small uncertainty. A few show a small, marginally significant, hardening. For the examples shown in Fig. 5, we also add in the lower panels of Fig. 5 the expected $\b$-evolutions from the dust scattering model. A few model parameters ($R$, $q$ and $s$) were set free to change and then were optimised in each example in order to best reproduce the observed plateau and post-plateau LC. The expected $\b$ is systematically larger than the observed one even at the beginning of the plateau phase, because the softening has already begun there. The expected strong evolution in $\b$ distinctly differs from the stableness of the observed $\b$ during the plateau.

To summarise this section, though the dust scattering model can nicely fit the LCs of plateau and post-plateau decay (see also Shao et al. 2008), the expected large value and strong evolution of the spectral index sharply contradict the data.

%%%%%%%%%%%%%%%%%%%%%%%%%%%%%%%%%Begin beta history figure%%%%%%%%%%%%%%%%%%%%%%%%%%%
\begin{figure*}
\centerline{\includegraphics[width=17.7cm,angle=0]{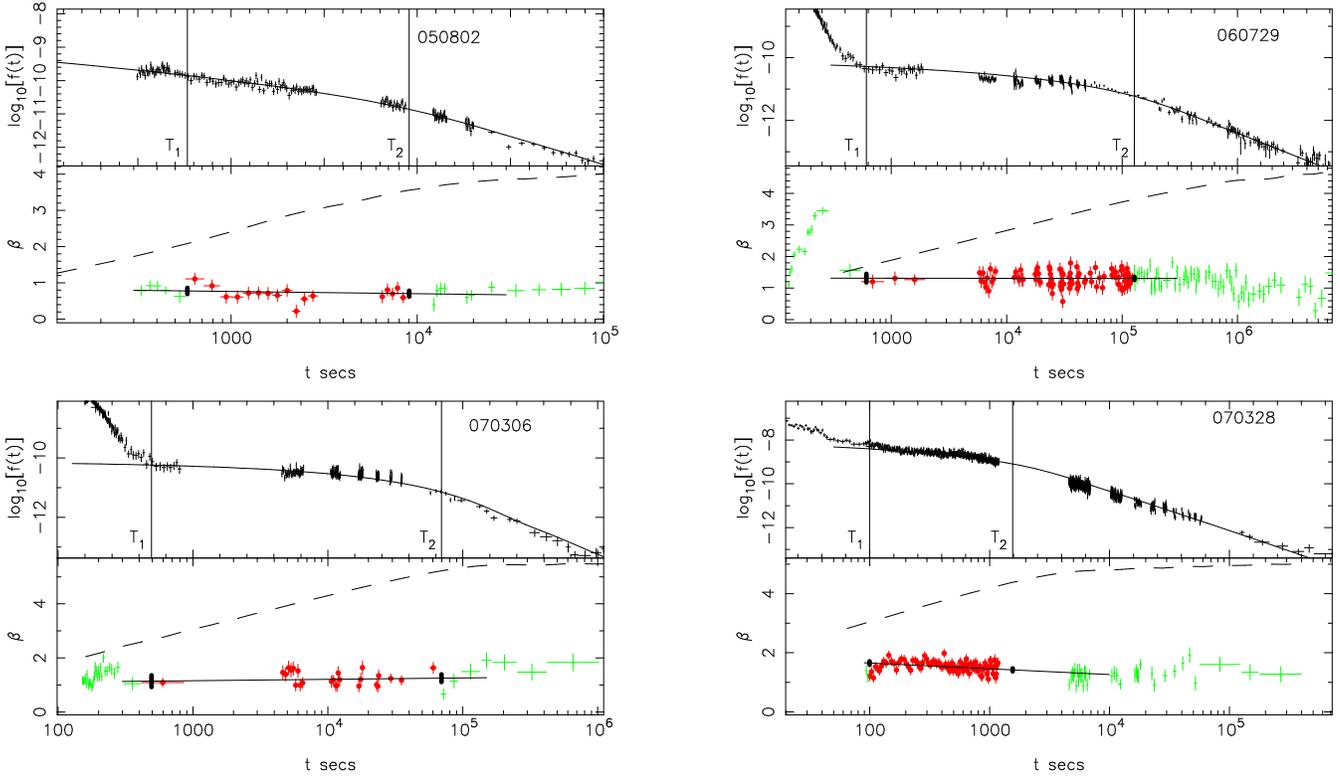}}
\caption{The X-ray LC and the time history of the spectral index $\b$ for four examples of GRBs with plateaus and the corresponding expectations from the dust scattering model. In the {\bf upper} panel for each example, $T_1$ and $T_2$ mark the beginning and the end of the plateau phase, respectively. The plateau phase and the post-plateau decay LC is mimicked by the dust scattered emission through adjusting the parameters ($R$, $q$ and $s$) of a single dust screen model (using an extended dust zone model does not change the result), shown as the \emph{solid} line. In the {\bf lower} panel for each example, the time-resolved $\b$ are plotted as filled circles inside the [$T_1$, $T_2$] window and as crosses outside the window. A linear function of $\log(t)$, as shown by the \emph{solid straight} line, is fitted to the time history of $\b$ within the [$T_1$, $T_2$] window. The fit extrapolation at $T_1$ and $T_2$ gives $\b_1$ and $\b_2$, respectively, as marked by the filled bars. The \emph{dashed} lines are the expected $\b$-evolutions for the dust models that were optimised in the upper panels.}
\end{figure*}
%%%%%%%%%%%%%%%%%%%%%%%%%%%%%%%End beta history figure%%%%%%%%%%%%%%%%%%%%%%%%%%%%%%%

%%%%%%%%%%%%%%%%%%%%%%%%%%%%%%%%%%Begin Tab. 1%%%%%%%%%%%%%%%%%%%%%%%%%%%%%%%%%%%
\begin{table*}
\caption{The spectral indices at the beginning and at the end of the X-ray plateaus for a sample of GRBs. From a larger sample of GRBs showing well defined plateau phases, only those with high signal-to-noise ratio and long time coverage in the plateau are selected. $T_1$ marks the transition of the LC from the prompt component to the plateau component. $T_2$ marks the end of the plateau and the transition to the final power-law decay. $\b_1$ and $\b_2$ are given by the extrapolation of a function fit to the evolution of the spectral index of available data within a window defined by $T_1$ and $T_2$. The error in $\b$ is at 90\% confidence level. }
\begin{center}
\begin{tabular}{llrlrr}
\hline
%%Entering the name for each column.
 GRB & $T_1$ (10$^2$s) & $\b_1$ &
 $T_2$ (10$^4$s) & $\b_2$ & $\chi^2/n_{dof}$\\
\hline \hline
050315 &   4.0 &    1.24 $\pm$ 0.17 & 2.48 &    0.99 $\pm$ 0.08         & 44.9/29\\
050319 &   6.6 &    0.95 $\pm$ 0.14  & 4.65 &   0.88 $\pm$ 0.13         & 20.6/17\\
050401 &   7.9 &    0.88 $\pm$ 0.06  &  0.75 &    0.90 $\pm$ 0.10       & 81.2/51\\
050713B &  7.2 &    0.87 $\pm$ 0.08  & 2.79 &    0.93 $\pm$ 0.15        & 16.9/19\\
050802  &  5.8 &    0.78 $\pm$ 0.09  &  0.90 &    0.70 $\pm$ 0.09       & 31.1/27\\
050803 &   2.9 &    0.76 $\pm$ 0.11  &   0.078 &    0.62 $\pm$ 0.21     & 9.11/9\\
  060306 &   4.6 &    1.34 $\pm$ 0.14  &  0.80 &    1.31 $\pm$ 0.15     & 20.9/12\\
  060502A &  4.6 &    1.26 $\pm$ 0.22  & 1.78 &    0.94 $\pm$ 0.17      & 9.4/7\\
  060510A &   0.92 &  1.03 $\pm$ 0.22  & 1.30 &    1.02 $\pm$ 0.09      & 26.2/23\\
  060607A &  8.4 &    0.96 $\pm$ 0.12  & 5.57 &    0.83 $\pm$ 0.13      & 28.5/38\\
  060614 &    22 &    1.03 $\pm$ 0.19  & 11.5 &    0.81 $\pm$ 0.11      & 27.8/22\\
  060729 &   6.1 &    1.31 $\pm$ 0.14  & 12.7 &    1.31 $\pm$ 0.06      & 143.8/114\\
  060813 &   0.32 &   1.03 $\pm$ 0.30  &   0.047 &    0.82 $\pm$ 0.11   & 19.5/19\\
  060814 &    12 &    0.69 $\pm$ 0.11  & 1.90 &    0.68 $\pm$ 0.09      & 46.4/31\\
  061121 &   2.0 &    1.12 $\pm$ 0.11   & 0.24 &    0.96 $\pm$ 0.10     & 20.3/27\\
  061222A &  1.6 &    1.36 $\pm$ 0.19  &  0.158 &    1.26 $\pm$ 0.10    & 49.9/26\\
  070129 &    14 &    1.58 $\pm$ 0.28  & 2.78 &    1.18 $\pm$ 0.14      & 14.0/12\\
  070306 &   4.9 &    1.15 $\pm$ 0.22 & 6.95 &    1.25 $\pm$ 0.14       & 43.4/28\\
  070328 &   1.0 &    1.65 $\pm$ 0.06  &  0.16 &    1.42 $\pm$ 0.05     & 151.4/107\\
  070420 &   2.1 &    1.27 $\pm$ 0.23  &  0.37 &    0.87 $\pm$ 0.12     & 16.9/15\\
  070508 &   0.69 &   0.77 $\pm$ 0.08  &  0.10 &    0.63 $\pm$ 0.05     & 107.1/93\\

\hline
\end{tabular}
\end{center}
\end{table*}
%%%%%%%%%%%%%%%%%%%%%%%%%%%%%%%%%End Tab. 1%%%%%%%%%%%%%%%%%%%%%%%%%%%%%%%%%%%

%%%%%%%%%%%%%%%%%%%%%%%%%%%%%%%%Begin Fig. 6%%%%%%%%%%%%%%%%%%%%%%%%%%%%%%%%%
\begin{figure}
\centerline{\includegraphics[width=6.35cm,angle=270]{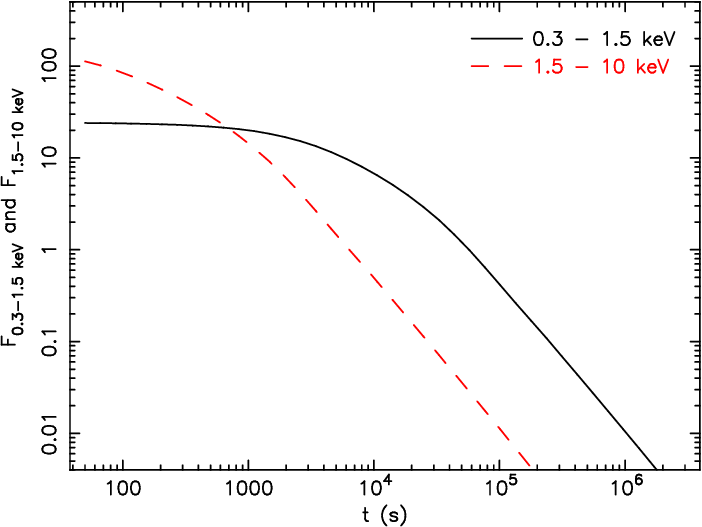}}
\hspace{0.04cm} \centerline{\includegraphics[width=6.45cm,angle=270]{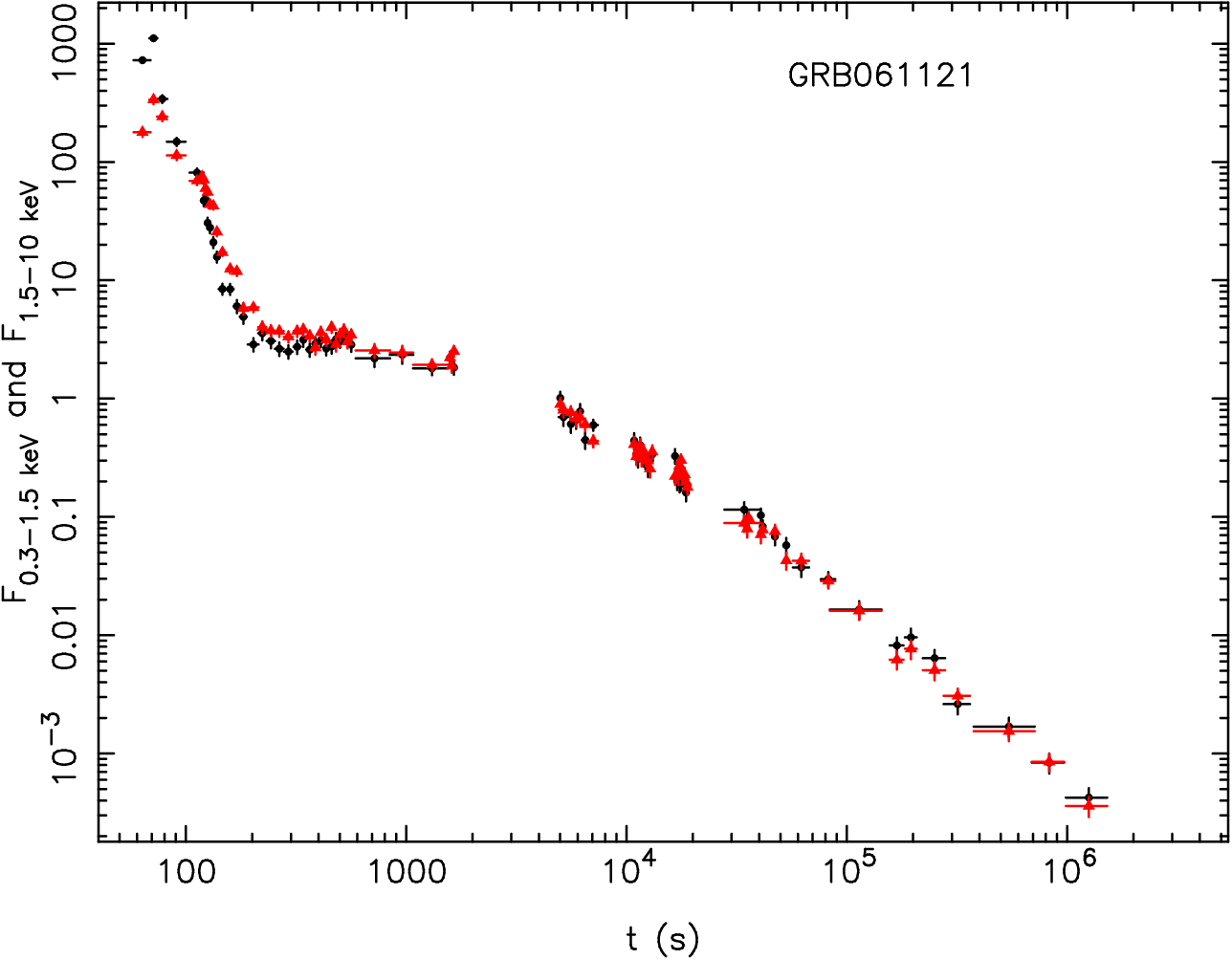}}
\caption{The LCs in the soft and the hard X-ray channels, respectively, as predicted by the dust echo model and observed in GRBs. {\bf Top}: The LCs of the echo emission from an extended dust zone model. The values of the model parameters $R_-$ and $R_+$ are chosen such that the echo emission 0.3 - 10 keV LC best mimics the plateau of GRB061121. Other model parameters are the same as the fiducial ones in Fig. 1. {\bf Bottom}: The LCs of a typical GRB `plateau' in 0.3 - 1.5 keV (\emph{squares}) and 1.5 - 10 keV (\emph{triangles}), respectively. The early rapid decay at $t< 200$ s of the observed LC does not belong to the plateau phase and it is thought to be of a different origin.}
\end{figure}
%%%%%%%%%%%%%%%%%%%%%%%%%%%%%%%%End Fig. 6%%%%%%%%%%%%%%%%%%%%%%%%%%%%%%%%%%%

%%%%%%%%%%%%%%%%%%%%%%%%%%%%%%%%%%%%%%%%%%%%%%%%%%%%%%%%%%%%%%%%%%%%%%%%%%%%%%%%%
\section{X-ray light curves in soft and hard energy channels}

Another prediction from the dust scattering model is the different temporal behaviours of the LCs in low and high energy channels. This can be seen from equation (4): observations carried out at photon energy $E$ should see the end of the plateau at $t_c \propto E^{-2}$. This feature is demonstrated in the top panel of Fig. 6 where we calculated the LCs of scattered emission from an extended dust zone in 0.3 - 1.5 keV and 1.5 - 10 keV, respectively. It shows that the soft X-ray LC has a more extended plateau than the hard X-ray one.

We find that for all bursts in our sample of plateau GRBs, the temporal behaviours in soft vs. hard X-ray LCs look identical. As an example, the bottom panel of Fig. 6 shows the soft vs. hard X-ray LCs of GRB 061121 which has a long, dense time coverage and the best photon statistics among all bursts in the sample (Page et al. 2007). The soft and hard X-ray LCs for this burst are identical. This feature rules out the dust scattering model for the plateau.

%%%%%%%%%%%%%%%%%%%%%%%%%%%%%%%%%%%%%%%%%%%%%%%%%%%%%%%%%%%%%%%%%%%%%%%%%%%%%%%%%%%%%
\section{Optical extinction due to the dust}

%%%%%%%%%%%%%%%%%%%%%%%%%%%%%%%Begin Tab. 2%%%%%%%%%%%%%%%%%%%%%%%%%%%%%%%%
\begin{table*}
\caption{The fiducial host rest frame optical depth at $E=$ 1 keV required by the dust scattering model and the associated rest frame visual extinction for a sub-sample of GRBs with well observed shallow X-ray decays. The fluence and spectral data are from Liang et al. (2007). $A_V^{(1)}$ is given via equation (7) and $A_V^{(2)}$ via equation (8). References to the GRB redshifts: 050315 - Kelson \& Berger (2005); 050319 - Jakobsson et al. (2006); 050401 - Fynbo et al. (2005); 050803 - Bloom et al. (2005); 060210 - Cucchiara et al. (2006); 060714 - Jakobsson et al. (2006); 060729 - Thoene et al. (2006); 060814 - Thoene et al. (2007). $z=2$ is assumed for GRBs without known $z$. }
\begin{center}
\begin{tabular}{lclclllll}
\hline
%%Entering the name for each column.
 \multirow{2}{*}{GRB} & $S_{\g}$ & \multirow{2}{*}{$\b_{\g}$} & $S_X$ &
 \multirow{2}{*}{$\b_X$} & \multirow{2}{*}{$z$} & \multirow{2}{*}{$\tau_0$} & \multirow{2}{*}{$A_V^{(1)}$} & \multirow{2}{*}{$A_V^{(2)}$}\\
 & (10$^{-7}$ erg cm$^{-2}$) &  & (10$^{-7}$ erg cm$^{-2}$) &
 \\
\hline \hline
050128 & 45 & 0.5 & 3.7 & 0.87 & & 25 & 414 & 162\\
050315 & 28 & 0.28 & 11 & 1.06 & 1.95 & 1.83 & 30 & 12\\
050319 & 8  & 0.25 & 1.3 & 1.00 & 3.24 & 8.4 & 140 & 56\\
050401 & 140 & 0.15 & 9.3 & 0.91 & 2.9 & 12 & 200 & 79\\
050713B & 82 & 0.0 & 3.3 & 0.85 & & 6.4 & 108 & 42\\
050803 & 39 & 0.05 & 6.0 & 0.76 & 0.42 & 0.5 & 8.4 & 3.4\\
050822 & 34 & 0.0 & 4.1 & 1.29 & & 1.7 & 29 & 12\\
060210 & 77 & 0.52 & 10 & 1.06 & 3.91 & 39 & 650 & 260\\
060714 & 30 & 0.99 & 1.5 & 1.15 & 2.71 & 385 & $6.5\times10^3$ & $2.6\times10^3$\\
060729 & 27 & 0.86 & 20 & 1.35 & 0.54 & 2.6 & 43 & 100\\
060813 & 55 & -0.47 & 7.3 & 1.09 &  & 0.22 & 3.6 & 1.4\\
060814 & 150 & 0.56 & 6.9 & 1.11 & 0.83 & 18 & 308 & 124\\
070129 & 31 & 1.05 & 1.5 & 1.25 & & 315 & $5\times10^3$ & $2\times10^3$\\
\hline
\end{tabular}
\end{center}
\end{table*}
%%%%%%%%%%%%%%%%%%%%%%%%%%%%%%%End Tab. 2%%%%%%%%%%%%%%%%%%%%%%%%%%%%%%%%%%

The dust grains which scatter the X-ray photons will also cause extinction in the optical band. This can provide an additional constraint for the dust scattering model. Thus we estimate the extinction in V Band, $A_V$, caused by the dust required for the model and compare it with $A_V$ derived directly from optical observations.

Predehl \& Schmitt (1995) found an empirical relation between the X-ray dust scattering optical depth $\tau(E)$ and $A_V$ for the X-ray halos of 24 galactic X-ray point sources:
\begin{equation}\label{predehl95}
\tau(E)= 0.06 A_V (E/1 {\rm keV})^{-2},
\end{equation}
where $\tau(E)$ is obtained from modelling the X-ray halo surface brightness distributions with dust grain properties similar to the fiducial ones we used in our calculations. Draine \& Bond (2004) derived a similar relation based on a dust model developed by Draine (2003):
\begin{equation}
\tau(E)= 0.15 A_V (E/1 {\rm keV})^{-1.8}.
\end{equation}
Note that Eq. (\ref{predehl95}) is for the dust in the Milky Way (MW), while for GRB hosts most absorption fits tend to favour the Small Magellanic Cloud (SMC) extinction law (e.g., Schady et al. 2007; but see the discussion below toward the end of this section). It was shown that the difference between the the MW and SMC extinction laws can be well reproduced in a model by adjusting the relative abundances of graphite and silicate grains, while leaving all other dust properties fixed; in this case $A_V$ and $\tau(E)$ at $E \ge 7$ eV are both the same for these two environments (Pei 1992). Thus it is viable to apply Eq. (\ref{predehl95}) to GRB hosts.

The above relations are for quantities in the rest frame of the source. If the source is at cosmological distances, like GRB hosts, we have to take into account the cosmological redshift of the photon energy when calculating $\tau(E)$ and $A_V$ from the observed quantities. The dust scattering optical depth, $\tau_0= \tau (E=1 {\rm keV})$, in the rest frame of the GRB host at a redshift $z$ can be estimated by
\begin{equation}
S_{X, 1} \approx (1+z)^{-2}  \tau_0 S_{\g, 1},
\end{equation}
where $S_{\g,1}= S_{\g}(E=1{\rm keV})$ is the specific fluence extrapolated from the Burst Alert Telescope (BAT) total fluence during the burst to 1 keV, $S_{X,1}= S_X(E= 1{\rm keV})$ is the specific fluence at 1 keV during the plateau phase, both measured in the observer's frame. There is a factor of $(1+z)^{-2}$ because the ratio $(S_{X, 1}/S_{\g, 1})$ is actually equal to the host rest-frame $\tau(E)$ at $E= (1+z)$ keV and $\tau(E) \propto E^{-2}$ (cf. Eq. \ref{tau_a_E}).

We select a sub-sample of GRBs which have good XRT temporal coverage during the afterglow phase from the sample of Liang et al. (2007) that provided $S_{\g}$ and $S_X$ for those bursts. Then $\tau_0$ is calculated and the associated $A_V$ is inferred from $\tau_0$ via Eq. (7) and (8). The sample and the results are listed in Table 2. Almost in all cases (except for two) $\tau_0$ is $>$ 1 and some even have $\tau_0 >$ 10, which means this model requires that only a very tiny fraction of photons with energy of 1 keV in the host rest frame can escape the dust without scattering. None of the sub-sample have $A_V < 1$ and 85\% of them have $A_V > 10$.

In comparison, Schady et al. (2007) determined $A_V$ for 6 GRB afterglows from the Ultraviolet and Optical Telescope (UVOT) to XRT Spectral Energy Distributions (SEDs), with $A_V$ ranging from 0.1 to 0.7. These extinctions are significantly smaller than that expected from the dust scattering model. Note that a considerable fraction (1/4 - 1/3) of the plateau X-ray afterglows have bright optical counterparts (Figure 2 of Liang et al. 2007).

%%%%%%%%%%%%%%%%%%%%%%%%%%%%%%%%Begin Fig. 7%%%%%%%%%%%%%%%%%%%%%%%%%%%%%%%%%
\begin{figure}
\centerline{\includegraphics[width=6.4cm,angle=270]{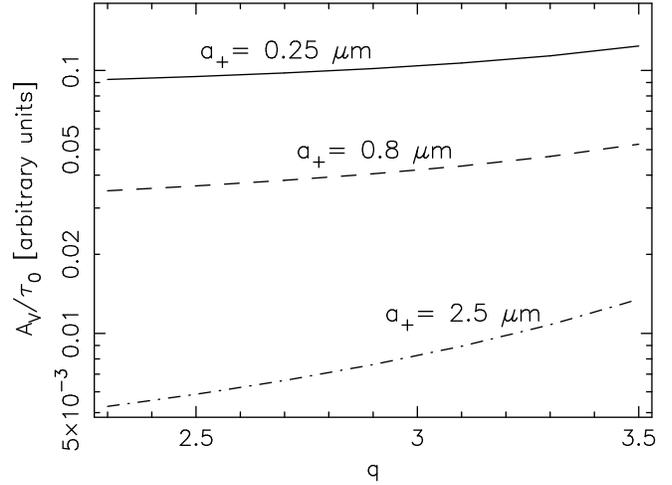}}
\caption{The dependence of the ratio $A_V/\tau_0$ of the dust on the grain size distribution parameters $a_+$ and $q$. The dust is assumed to have a grain size distribution $dN(a)/da \propto a^{-q}$ within the grain size range $[a_-, a_+]$, where $N(a)$ is the column density of all grains with size $\le a$.}
\end{figure}
%%%%%%%%%%%%%%%%%%%%%%%%%%%%%%%%End Fig. 7%%%%%%%%%%%%%%%%%%%%%%%%%%%%%%%%%%%

But a cautiousness has to be taken regarding the $A_V$ determinations above. There are some recent studies that show the dust
properties of GRB hosts does not resemble those of any galaxy
in our neighborhood. In particular, Chen et al. (2006), Perley
et al. (2008) and Li et al. (2008) have found for some GRBs the
modeled extinction curve is ``gray'', i. e., much flatter than
any of the templates (MW, SMC, the Large Magellanic Cloud). Since our determination of $A_V$
has used the $A_V$ - $\tau_0$ relations appropriate for these template-type
dust, these findings are likely to raise uncertainties in the determined $A_V$.

The dust grain size distribution is usually
described by a power law with an index $q$ and within a range of
grain size ($a_-$, $a_+$). A ``gray'' extinction curve could be
due to a flatter grain size distribution (smaller $q$) or a larger $a_+$, as
suggested by Li et al. (2008). We calculated $A_V$ and $\tau_0$ independently for a dust model with the composition resembling the SMC and the ``gray''
type that these authors have found for some GRBs, to see the dependence of their ratio on the grain size distribution parameters. The $A_V$ is calculated by the
following equation (Weingartner \& Draine 2001)
\begin{equation}
A_V= (2.5\pi \log{e}) \int_{a_-}^{a_+} a^2 Q_{ext}(a, \lambda_V) \frac{dN(a)}{da} da,
\end{equation}
where $Q_{ext}$ is the extinction efficiency factor, usually a function of grain size and photon energy. Draine \& Lee (1984) calculated $Q_{ext}(a, \lambda)$ for graphite and silicate grains. Since the GRB dust environments are described either by SMC or ``gray'' type extinction curve (Shady et al. 2007; Li et al. 2008), for both of which a good dust composition model needs silicate only (Pei 1992; Li et al. 2008), we use $Q_{ext}(a, \lambda)$ for silicate only. The dust scattering optical depth can be calculated by $\tau_0=\int_{a_-}^{a_+} \tau_a(E=1{\rm keV}) da$, where $\tau_a(E)$ is given by Eq. (\ref{tau_a_E}). Note that the normalizations of $A_V$ and $\tau_0$ are unimportant here because we are looking at only their dependences on $a_+$ and $q$.

The ratio of $A_V /
\tau_0$ for varying $a_+$ and $q$ is shown in Fig. 7. We find that $A_V / \tau_0$ is only slightly dependent on $q$ -- $A_V / \tau_0$
decreases by factor of $\sim 2$ for $q$ changing from 3.5 to 2.6; but
$A_V / \tau_0$ is more sensitive to $a_+$ -- it decreases by a
factor of $\sim 10$ when $a_+$ changes from 0.25 $\mu m$ to 2.5
$\mu m$. Even after taking these effects into account, the $A_V$ we
obtained for our GRB sample are still very large (for one
half of the sample, $A_V \gtrsim 10$). One of the real problems with
the dust scattering model for the X-ray plateau phase is
that it requires $\tau_0 \geq 10$ for one half of our sample (see Table 2),
which seems physically unreasonable.

Thus, the dust scattering model is not a viable explanation for the X-ray plateaus because of the large extinction in the optical band it predicts but not observed.

%%%%%%%%%%%%%%%%%%%%%%%%%%%%%%%%%%%%%%%%%%%%%%%%%%%%%%%%%%%%%%%%%%%%%%%%%%%%
\section{Conclusion and Discussion}

We have shown that in the dust scattering model the scattered X-ray emission must experience strong softening spectral evolution, with a significant change of the spectral index in 0.3 - 10 keV of $\D \b \sim 2 - 3$ from the emerging of the plateau to its end. However, for a sample of GRBs with X-ray plateaus and with good quality data, no softening spectral evolution during the plateau phase is found, and in a few cases even traces of slight hardening are seen.

The change of $\b$ according to the model does not depend on the spectral index of the source emission. The Rayleigh - Gans approximation is used in this paper to calculate the scattering cross section of the dust grain. It was claimed that this approximation tends to overestimate the scattering efficiency below 1 keV, typically by a factor of 4 at 0.5 keV and a factor of 2 at 1 keV, mainly due to the absorption of the soft X-ray photons by the K and L shell electrons in the dust grain (Smith \& Dwek 1998), and that could change the spectral slope at the soft end ($< 1$ keV) and counteract against the softening (Shao et al. 2008). But we argue that this effect dose not alleviate the expected softening, because the discrepancy between the real scattering cross section and the Rayleigh - Gans approximation caused by this effect, which is mainly below 1 keV, must have been largely accounted for by the required neutral $H$ absorption in the routine power-law fit to the plateau spectra. The XRT spectral index is mainly determined by the photons with energy above 1 keV which is not affected by this effect. Moreover, this effect is time independent while the softening we consider is a strongly time dependent behaviour. Dust destruction by the GRB prompt emission is of very little relevance here because it happens within a distance smaller than the location of the dust considered in this work (e.g., Waxman \& Draine 2000). Thus the Rayleigh - Gans approximation is sufficiently accurate for the effect considered in this work.

The dust scattering model also predicts very different temporal behaviours in the soft X-ray vs. hard X-ray LCs; the plateau lasts longer in soft X-rays. But this feature is not found in the data. Furthermore, the large scattering optical depth of the dust required by this model in order to explain the X-ray plateaus leads to extremely large extinction in optical - $A_V \gtrsim 10$. This is inconsistent with the observed extinctions for GRBs.

We conclude that the dust scattering model, though very attractive, can not explain the X-ray plateaus seen in most GRB afterglows. Although it is very likely that dust exists near the site of GRBs, and will scatter some fraction of the prompt and afterglow X-rays, this scattered emission is not a dominant contributor to the observed X-ray plateau. For those cases where an achromatic break at the end of the plateau is seen, a late, steady energy injection to the external shock is a more likely mechanism for producing the observed X-ray plateau, though it may not be able to work well for the cases with chromatic breaks.

%%%%%%%%%%%%%%%%%%%%%%%%%%%%%%%%%%%%%%%%%%%%%%%%%%%%%%%%%%%%%%%%%%%%%%%%%%%%%%
\section*{Acknowledgement}

We thank Alin Panaitescu for useful discussions and constructive suggestions, and Zi-Gao Dai, Lang Shao for useful comments on this paper, also the referee for helpful comments. R-FS thanks Rodolfo Barniol Duran for carefully reading the manuscript and providing comments. PK and R-FS were supported in part by a NSF grant AST-0406878. PAE acknowledges support from the UK Science and Technology Facilities Council (STFC). This work made use of data supplied by the UK Swift Science Data Centre at the University of Leicester funded by STFC.

%%%%%%%%%%%%%%%%%%%%%%%%%%%%%%%%%%%%%%%%%%%%%%%%%%%%%%%%%%%%%%%%%%%%%%%%%%%%%%%%

%%%%%%%%%%%%%%%%%%%%%%%%%%%%%%%%Begin Reference%%%%%%%%%%%%%%%%%%%%%%%%%%%%%%%%%

\end{document}